\documentclass[12pt]{article}
\usepackage{amssymb,amsmath,epsfig}

\begin{document}

\title{\bf Dynamics of Warm Chaplygin Gas Inflationary Models With Quartic Potential }
\author{Abdul Jawad$^a$ \thanks {abduljawad@ciitlahore.edu.pk, jawadab181@yahoo.com}, Sadaf Butt$^b$
and Shamaila Rani$^a$ \thanks {shamailatoor.math@yahoo.com;
drshamailarani@ciitlahore.edu.pk}\\
$^a$ Department of Mathematics, COMSATS Institute of\\
Information
Technology, Lahore, Pakistan.\\
$^b$ Department of Mathematics, Lahore Leads University,\\ Lahore,
Pakistan.}

\date{}

\maketitle
\begin{abstract}
Warm inflationary universe models in the context of generalized
chaplygin gas, modified chaplygin gas, generalized cosmic chaplygin
gas are being studied. The dissipative coefficient of the form
$\Gamma\propto T$, weak and strong dissipative regimes are being
considered. We use quartic potential
$\frac{\lambda_{*}\phi^{4}}{4}$, which is ruled out by current data
in cold inflation but in our models it is analyzed that it is in
agreement with the WMAP$9$ and latest Planck data. In these
scenarios, the power spectrum, spectral index, and tensor to scalar
ratio are being examined under the slow roll approximation. We show
the dependence of tensor scalar ratio $r$ on spectral index $n_{s}$
and observe that the range of tensor scalar ratio is $r<0.05$ in
generalized chaplygin gas, $r<0.15$ in modified chaplygin gas, and
$r<0.12$ in generalized cosmic chaplygin gas models. Our results are
in agreement with recent observational data like WMAP$9$ and latest
Planck data.
\end{abstract}

\section{Introduction}

It is well known that inflation presents most compelling solution of
many problems of big bang model, namely the horizon, flatness,
homogeneity and monopoles problems \cite{1}. The most fascinating
feature of inflationary universe model is that it interprets the
origin of observed anisotropy in the cosmic microwave background
radiations, and also the distribution of large scale structures
\cite{2}. But some questions arises in theory of inflation, one of
them is how to end this inflationary epoch and enter in bing bang
phase. Warm inflation provides a possible solution to this problem.
Standard inflation known as cold inflation, has two regimes slow
roll and reheating. In slow roll limits, universe expands as
potential energy dominates the kinetic energy and interaction of
inflation (scalar field) with other fields become negligible. In
reheating epoch, kinetic energy is comparable to potential energy
and inflation oscillates around the minimum of its potential while
losing its energy to massless particles. After reheating, the
universe is filled with radiation.

Warm inflation provides a mechanism in which reheating is avoided.
During the warm inflationary period,  dissipative effects are
important, so that radiation production takes place at the same time
as inflationary expansion. A strong regime in warm inflation is that
in which damping effects on inflation dynamics of radiation field
are strong, these dissipating effects originates from a friction
term which describe the physical process of decay of inflation field
into a thermal bath due to its interaction with other field. Decay
of remaining inflationary field or dominant radiation create the
mater component of universe. Warm inflation come to an end when
universe heats up and become radiation dominated and gets connected
with the big bang scenario \cite{3,3t}. In standard inflation
density perturbations are generated due to quantum fluctuations
associated to the inflation scalar field, which are necessary for
the large scale structure formation at the late time in the
evolution of the universe. However, in warm inflation, thermal
fluctuations instead of quantum fluctuations become a source of
density perturbations \cite{36,37}.

Monerat et al. \cite{11} studied cosmology of the early universe and
the initial condition for inflation in a model with radiation and
chaplygin gas (CG). Antonella et al. \cite{4t} discussed warm
inflation on brane. Del campo and Herrera \cite{9} considered warm
inflationary model with generalized chaplygin gas (GCG), they used a
standard scalar field and dissipation coefficient of the form
$\Gamma\propto \phi^{n}$ and then develop the model with chaotic
potential. Setare and Kamali investigated warm tachyon inflation by
assuming intermediate \cite{2t} and logamediate scenario \cite{1t}.
Bastero-Gill et al. obtained the expressions for the dissipation
coefficient in supersymmetric (SUSY) models in \cite{6}. This result
provides possibilities for realization of warm inflation in SUSY
field theories.

Herrera et al. \cite{9t} studied intermediate inflation in the
context of GCG using standard and tachyon scalar field. Same authors
also dealt with dissipation coefficient
$\Gamma=c\frac{T^{m}}{\phi^{m-1}}$ in the context of warm
intermediate and logamediate inflationary models \cite{10}. They
also studied warm inflation in loop quantum cosmology with the same
dissipative coefficient \cite{10t}. Bastero-Gill et al. in \cite{5}
have also explored inflation by assuming the quartic potential.
Sharif and Saleem \cite{8} studied inflationary models with
generalized cosmic chaplygin gas (GCCG). Setare and Kamali analyzed
warm viscous inflation on brane in \cite{3t}. They have also
considered a generalized de-sitter scale factor including single
scalar field and studied q-inflation in the context of warm
inflation with two forms of damping term \cite{7}. Panotopoulos and
Videla \cite{4} investigated the quartic potential model in the
framework of warm inflation by using a decay rate proportional to
temperature and showed that it is compatible with latest
observational data. We extend this work with the inclusion of
chaplygin gas (CG) models.

The goal of present work is to investigate the realization of warm
quartic inflationary model in the context of CG models. This paper
is organized as follows: Next section deals with basic background
equations of warm inflationary scenario. In section \textbf{3}, we
construct models with (GCG), (MCG) and (GCCG) by using a quartic
potential. In the last section, we summarized our results.

\section{Basic Inflationary Scenario}

We start by using Friedmann Robertson Walker (FRW) metric and
consider a spatially flat universe which contains a self interacting
inflation field $\phi$ and radiation field, where $V(\phi)$ is
scalar potential, $\rho_{\phi}$ and $\rho_{\gamma}$ are energy
densities of inflation field and radiation field respectively, then
write down a modified Friedmann equation of the form
\begin{equation}\label{1}
H^{2}=\frac{1}{3M_{p}^{2}}(\rho_{\phi}+\rho_{\gamma}),
\end{equation}
where $M_{p}=\frac{1}{\sqrt{8\pi G}}$ is reduced planck mass, and
$\rho_{\phi}=\frac{\dot{\phi}^{2}}{2}+V(\phi)$,
$P_{\phi}=\frac{\dot{\phi}^{2}}{2}-V(\phi)$ are the energy densities
and potential of scalar field, respectively. Energy-momentum
conservation leads to the following equations \cite{3,3t}
\begin{eqnarray}\label{4}
&&\dot{\rho_{\phi}}+3H(\rho_{\phi}+P_{\phi})= -
\Gamma\dot{\phi}^{2},\quad \dot{\rho_{\gamma}}+4H(\rho_{\gamma})=
 \Gamma\dot{\phi}^{2}.
\end{eqnarray}
Dynamics of  warm
inflation is described by adding a friction term in the equation of
motion given by
\begin{eqnarray}\label{2}
\ddot{\phi}+(3H+\Gamma)\dot{\phi}+V'=0,
\end{eqnarray}
$\Gamma$ is the dissipation coefficient. During the inflation era,
$\Gamma$ is responsible for the decay of scalar field into
radiation, this decay rate can be a  function of scalar field or
temperature or depend on both $\Gamma(T,\phi)$ or simply a constant.
During warm inflation production of radiation is quasi-stable, i.e.
$\dot{\rho_{\gamma}}\ll4H\rho_{\gamma}$ and  $\dot{\rho_{\gamma}}\ll
\Gamma\dot{\phi}^{2}$ \cite{3,3t,36,41,42,43}, the energy density
associated with scalar field dominates over the energy density of
radiation field, i.e. $\rho_{\phi}\gg\rho_{\gamma}$. Assuming the
set of slow roll conditions, i.e. $ \dot{\phi}^{2}\ll V(\phi)$ and
$\ddot{\phi}\ll(3H+\Gamma)\dot{\phi}$ \cite{3,3t}, then the
equations of motion reduces to
\begin{equation}\label{5}
3H(1+R)\dot{\phi}\simeq-V',\quad 4H\rho_{\gamma}\simeq
\Gamma\dot{\phi}^{2},
\end{equation}
here dot mean derivatives with respect to time and $
V'=\frac{\partial V}{\partial\phi} $. A dissipation coefficient is
basic quantity, which has been calculated from first principles in
the context of supersymmetry. In these models, there is a scalar
field with multiplets of heavy and light fields that make it
possible to obtain several expression for dissipation coefficient. A
general form for $\Gamma$ can be written as \cite{12,15}
\begin{equation}\nonumber
\Gamma=b\frac{T^{m}}{\phi^{m-1}},
\end{equation}
where $b$ is associated to dissipative microscopic dynamics and
exponent $m$ is integer. In literature different cases have been
studied for the different values of $m$, in special case $m=1$, i.e.
$\Gamma\propto T$ represent high temperature SUSY case, for the
value $m=0$ i.e. $\Gamma\propto \phi$ corresponds to an
exponentially decaying propagator in the high temperature SUSY
model, for $m=-1$ i.e. $\Gamma\propto \frac{\phi^{2}}{T}$, we have
agreement with non-SUSY case \cite{13,14}. We introduce a parameter
$R = \frac{\Gamma}{3H}$ which is the relative strength of thermal
damping compared to the expansion damping. In warm inflation,  we
can assume two possible scenarios, one is weak dissipative regime
defined as $R \ll 1$, in which Hubble damping is still the dominant
term, and the other one is strong dissipative regime defined as $R
\gg 1$, the $\Gamma$ controls damped evolution of the inflation
field in it.

Moreover, the thermalization energy density of radiation field can
be written as $\rho_{\gamma}=CT^{4}$, where constant
$C=\pi^{2}g_{*}/30$, and $g_{*}$ denotes the number of relativistic
degrees of freedom, in a Minimal Supersymmetric Standard Model
(MSSM), $g_{*}=228.75$ and $C\simeq 70$ \cite{36}. Using
Eq.(\ref{5}) and $\rho_{\gamma}\propto T^{4}$ temperature becomes
\begin{equation}\label{8}
T=\bigg[\frac{\Gamma
V'^{2}}{6^{2}CH^{3}(1+R)^{2}}\bigg]^{\frac{1}{4}}.
\end{equation}
Slow roll parameters of warm inflation are given by \cite{36}
\begin{eqnarray}\nonumber
\epsilon = \frac{-\dot{H}}{H^{2}},\quad
\eta=\frac{-\ddot{H}}{H\dot{H}},\quad
\beta=-\frac{1}{H}\frac{d}{dt}(ln\Gamma).
\end{eqnarray}
In warm inflation, slow roll conditions are expressed as  $\epsilon
\ll 1+R$, $\eta \ll 1+R$, $\beta \ll 1+R$. On the other hand, the
number of e-folds is calculated by using the standard formula
\begin{eqnarray}\label{1n}
N=\int_{t_{*}}^{t_{end}}H dt.
\end{eqnarray}
Here, $t_{*}$ and $t_{end}$ denotes the time when inflation starts
and comes to an end respectively.

Next, we discuss the perturbation parameters for the current
scenario by assuming CG models. The perturbation parameters of the
warm inflation are obtained in \cite{36}. The amplitude of the power
spectrum of the curvature perturbation is given by
\begin{equation}\label{9}
P_{R}=\bigg(\frac{\pi}{4}\bigg)^{\frac{1}{2}}\frac{H^{\frac{5}{2}}\Gamma^{\frac{1}{2}}T}{\dot{\phi}^{2}},
\end{equation}
we can calculate the scalar spectral index $n_{s}$ \textbf{by} using
$n_{s} = 1+\frac{dP_{R}}{dlnk}$ which is equivalent to
\begin{eqnarray}\label{10}
n_{s}=1-\frac{9\epsilon}{4}+\frac{3\eta}{2}-\frac{9\beta}{4}.
\end{eqnarray}
However tensor to scalar ratio turns out to be \cite{7}
\begin{equation}\label{12}
r=\frac{32G\dot{\phi}^{2}}{\Gamma^{\frac{1}{2}}\pi^{\frac{3}{2}}TH^{\frac{1}{2}}}.
\end{equation}
In the following, we take a standard scalar field and $\Gamma\propto
T$ to study how these conditions effects the inflationary dynamics
for quartic potential.

\section{Chaplygin Inflationary Models With Quartic Potential}

We consider a quartic potential
$V(\phi)=\frac{\lambda_{*}\phi^{4}}{4}$ which is a simple Higgs
potential developed in particle physics theories \cite{5t}. In the
following work, we assume an inflation decay rate $\Gamma=bT$ and
quartic potential in warm inflation models with chaplygin gas.

\subsection{Generalized Chaplygin Gas}

The CG is considered to be an alternative description of
accelerating expansion and it has a connection with string theory.
CG emerges as an effective fluid of generalized D-brane in a
$(d+1,1)$ space time where the action can be written as a
generalized born-infield action \cite{8x}. Kammshchick \cite{9x}
considered FRW universe composed of CG and showed that universe is
in agreement with current observation of cosmic acceleration. Its
extended form is GCG whose equation of state (EoS) is as follows
\begin{equation}\nonumber
P_{gcg}= -
\frac{A}{\rho_{gcg}^{\lambda}}  ,
\end{equation}
where $P_{gcg}$ and $\rho_{gcg}$ denote the pressure and energy
density respectively and $0<\lambda\leq1$, and A is the positive
constant. The energy density of GCG can be obtained by using
equation of continuity and given by
\begin{equation}\label{1x}
\rho_{gcg}=\bigg(A+\frac{B}{a^{3(1+\lambda)}}\bigg)^{\frac{1}{1+\lambda}}
,
\end{equation}
where $B$ is a positive integration constant and $a$ is scale
factor. We start with the modified Friedmann equation of the form
\begin{equation}\label{14}
H^{2}=\frac{1}{3M_{p}^{2}}\bigg(\big(A+\rho_{\phi}^{1+\lambda}\big)^{\frac{1}{1+\lambda}}+\rho_{\gamma}\bigg).
\end{equation}
This modification is possible due to an extrapolation of
Eq.(\ref{1x}) so that
\begin{eqnarray}\label{15}
\rho_{gcg}=\big(A+\rho_{m}^{1+\lambda}\big)^{\frac{1}{1+\lambda}}
\rightarrow\big(A+\rho_{\phi}^{1+\lambda}\big)^{\frac{1}{1+\lambda}},
\end{eqnarray}
where $\rho_{m}$ denotes the matter energy density. During the
inflation era,  energy density of scalar field dominates the energy
density of radiation field, i.e., $ \rho_{\phi}\gg\rho_{\gamma}$,
and it is of the order of potential i.e. $\rho_{\phi}\sim V$. For
simplicity, we take $\lambda=1$ for which the Friedmann equation
takes the form
\begin{eqnarray}\label{16}
H^{2}=\frac{1}{3M_{p}^{2}}\sqrt{A+\rho_{\phi}^{2}} \sim
\frac{1}{3M_{p}^{2}}\sqrt{A+V^{2}}.
\end{eqnarray}

\subsubsection{Weak Dissipative Regime}

Here, we consider weak dissipative regime where $R \ll 1$, the
Friedmann and Klein-Gordon equations take the standard form under
slow-roll approximation. By taking $\Gamma=bT$, the temperature of
radiation field becomes
\begin{eqnarray}\nonumber
T=\bigg(\frac{bV'^{2}}{6^{2}CH^{3}}\bigg)^{\frac{1}{3}}.
\end{eqnarray}
For weak dissipative regime, the slow roll parameters are as follows
\begin{eqnarray}\nonumber
\epsilon & =& \frac{M_{p}^{2}VV'^{2}}{2(A+V^{2})^{\frac{3}{2}}}
,\quad \eta
=\frac{M_{p}^{2}}{(A+V^{2})^{\frac{1}{2}}}\bigg(V''+\frac{V'^{2}}{V}-\frac{3VV'^{2}}{2(A+V^{2})}\bigg)
,\\\nonumber\beta &=&
M_{p}^{2}\bigg(\frac{4V''(A+V^{2})-3V'^{2}V}{6(A+V^{2})^{\frac{3}{2}}}\bigg).
\end{eqnarray}
By using the Eq.(\ref{1n}), the number of e-folds become
\begin{eqnarray}\nonumber
N=\frac{1}{M_{p}^{2}}\int_{\phi_{end}}^{\phi_{*}}\frac{\sqrt{A+V^{2}}}{V'}d\phi.
\end{eqnarray}
Amplitude of power spectrum given in Eq.(\ref{9}) takes the form
\begin{eqnarray}\label{18}
P_{R}&=&\bigg(\frac{81\pi
b^{2}}{12^{2}CV'^{2}}\bigg)^{\frac{1}{2}}\bigg(\frac{\sqrt{A+V^{2}}}{3M_{p}^{2}}\bigg)^{\frac{3}{2}}
.
\end{eqnarray}
By inserting the values  in Eq.(\ref{10}), scalar spectral index
turns out to be
\begin{eqnarray}
n_{s}-1&=&\frac{3M_{p}^{2}}{2(A+V^{2})^{\frac{1}{2}}}\bigg(\frac{-9VV'^{2}}{4(A+V^{2})}
-\frac{3}{2}\bigg(\frac{4V''(A+V^{2})-3V'^{2}V}{6(A+V^{2})}\bigg)\\\nonumber&&+V''+\frac{V'^{2}}{V}
\bigg).
\end{eqnarray}
The tensor to scalar ratio given in Eq.(\ref{12}) become
\begin{equation}
r=\frac{192G\sqrt{3CM_{p}^{2}}V'}{9b\pi^{\frac{3}{2}}(A+V^{2})^{\frac{1}{4}}}.
\end{equation}
We use $V=\frac{\lambda_{*}\phi^{4}}{4}$ and
$V'=\lambda_{*}\phi^{3}$ to express $r$ and $n_{s}$ as function of
$\phi$,
\begin{eqnarray}\nonumber
n_{s}-1&=&\frac{12M_{p}^{2}}{2(16A+\lambda^{2}_{*}\phi^{8})^{\frac{1}{2}}}
\bigg(\frac{-6\lambda^{3}_{*}\phi^{10}}{16A+\lambda^{2}_{*}\phi^{8}}
+4\lambda_{*}\phi^{2} \bigg),\quad
r=\frac{384G\sqrt{3CM_{p}^{2}}\lambda_{*}
\phi^{3}}{9b\pi^{\frac{3}{2}}(16A+\lambda^{2}_{*}\phi^{8})^{\frac{1}{4}}}.
\end{eqnarray}
\begin{figure} \centering
\epsfig{file=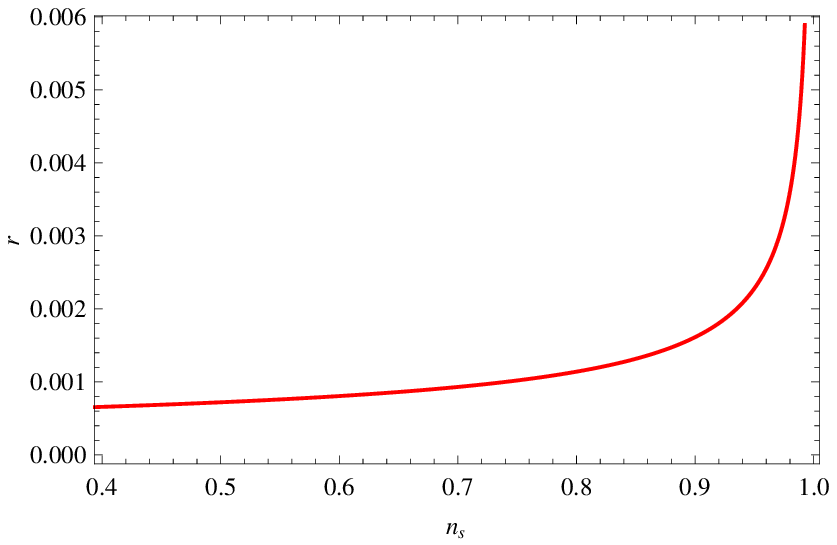,width=.45\linewidth}
\epsfig{file=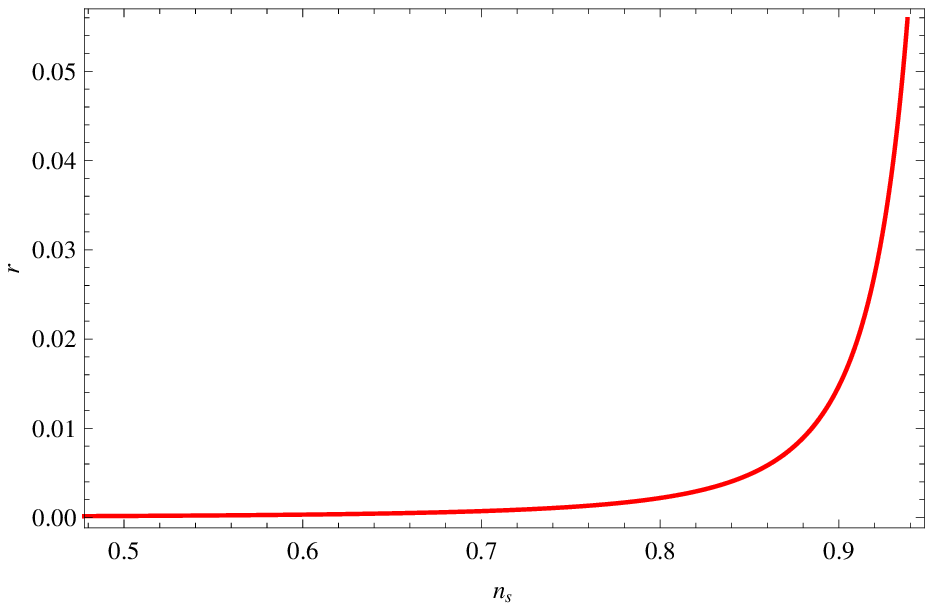,width=.45\linewidth}\caption{Plot of
tensor-scalar ratio $r$ versus scalar spectral index $n_{s}$ for GCG
model in weak dissipative regime (left panel) and strong dissipative
regime (right panel)}
\end{figure}

\subsubsection{Strong Dissipative Regime}

Now we consider a strong dissipative regime where $R \gg 1$, under
the slow roll approximation, the  temperature is given as
\begin{eqnarray}\nonumber
T=\bigg(\frac{V'^{2}}{4bCH}\bigg)^{\frac{1}{5}}.
\end{eqnarray}
For strong regime, the slow roll parameters leads to
\begin{eqnarray}\nonumber
\epsilon & =& \frac{M_{p}^{2}VV'^{2}}{2R(A+V^{2})^{\frac{3}{2}}}
,\\\nonumber\eta
&=&\frac{M_{p}^{2}}{R(A+V^{2})^{\frac{1}{2}}}\bigg(V''+\frac{V'^{2}}{V}-\frac{3VV'^{2}}{2(A+V^{2})}\bigg)
,\\\nonumber\beta &=&
\frac{1}{R}M_{p}^{2}\bigg(\frac{2V''(A+V^{2})-V'^{2}V}{5(A+V^{2})^{\frac{3}{2}}}\bigg).
\end{eqnarray}
The expression of number of e-folds takes the following form
\begin{eqnarray}\nonumber
N=\frac{1}{M_{p}^{2}}\int_{\phi_{end}}^{\phi_{*}}\frac{\sqrt{A+V^{2}}}{V'}Rd\phi.
\end{eqnarray}
In the similar way, we can obtain the amplitude of power spectrum,
scalar spectral index and tensor to scalar ratio as follows
\begin{eqnarray}\label{18}
P_{R}&=&\big(\frac{\pi}{4}\big)^{\frac{1}{2}}\frac{b^{\frac{9}{5}}}{V'^{\frac{3}{5}}(4C)^{\frac{7}{10}}}\bigg(\frac{\sqrt{A+V^{2}}}{3M_{p}^{2}}\bigg)^{\frac{9}{10}}
,\\\label{21}
n_{s}-1&=&\bigg(\frac{3(4C)^{\frac{1}{5}}}{b^{\frac{4}{5}}V'^{\frac{2}{5}}}\bigg)
\frac{(3M_{p}^{2})^{\frac{2}{5}}}{2(A+V^{2})^{\frac{1}{5}}}\bigg(V''+\frac{V'^{2}}{V}
-\frac{9VV'^{2}}{4(A+V^{2})}\\\nonumber&&-\frac{3}{2}\bigg(\frac{2V''(A+V^{2})-V'^{2}V}{5(A+V^{2})}\bigg)\bigg),
\\\label{22} r&=&
\frac{32G(4C)^{\frac{7}{10}}V'^{\frac{3}{5}}}{b^{\frac{9}{5}}\pi^{\frac{3}{2}}}\bigg(\frac{\sqrt{A+V^{2}}}
{3M_{p}^{2}}\bigg)^{\frac{1}{10}}.
\end{eqnarray}
In terms of scalar field, $r$ and $n_{s}$ turn out to be
\begin{eqnarray}\label{19}
n_{s}-1&=&\bigg(\frac{3(4C)^{\frac{1}{5}}}{b^{\frac{4}{5}}(\lambda_{*}\phi^{3})^{\frac{2}{5}}}\bigg)
\frac{(3M_{p}^{2})^{\frac{2}{5}}}{10(16A+\lambda^{2}_{*}\phi^{8})^{\frac{1}{5}}}
\bigg(\frac{-39\lambda^{3}_{*}\phi^{10}}{16A+\lambda^{2}_{*}\phi^{8}}
+26\lambda_{*}\phi^{2} \bigg),
\\\label{22}
r&=&\frac{32G(4C)^{\frac{7}{10}}(\lambda_{*}\phi^{3})^{\frac{3}{5}}}{b^{\frac{9}{5}}
\pi^{\frac{3}{2}}}\bigg(\frac{\sqrt{16A+\lambda_{*}^{2}\phi^{8}}}{12M_{p}^{2}}\bigg)^{\frac{1}{10}}.
\end{eqnarray}
We plot the $r$ versus $n_{s}$ for GCG models in Figure \textbf{1}
for weak (left panel) and strong dissipative regimes (right panel),
respectively. However, the parameters appearing in the model have
following values $M_{p}=1$, $\lambda_{*}=10^{-10}$, $A=10^{-45}$,
$b=0.3$. The trajectories in the Figure \textbf{1} show the
increasing behavior of $r$ with respect to $n_{s}$. It can be noted
that in weak dissipative regime (left panel of Figure \textbf{1})
that the range of tensor-to-scalar ration becomes $r<0.006$ for
$0.4<n_{s}<1$. However, it is $r=0.05$ corresponding to $n_{s}=0.96$
for strong dissipative regime (left panel of Figure \textbf{1}). It
is observed that WMAP$9$ \cite{17x} provides the value of tensor
scalar ratio as $r<0.13$ and spectral index is measured to be
$n_{s}=0.972\pm 0.013$. According to Planck data, $r<0.11$ and
$n_{s}=0.968\pm 0.006$ \cite{25}. In view of these observations, our
results for GCG model are compatible with observational data
\cite{17x,25}.

\subsection{Modified Chaplygin Gas}

The MCG has a equation of state as follows \cite{10x}
\begin{equation}\nonumber
P_{mcg}=\mu\rho_{mcg} - \frac{\nu}{\rho_{mcg}^{\lambda}}  ,
\end{equation}
where $P_{mcg}$ and $\rho_{mcg}$ denote the pressure and energy
density respectively and $0\leq\lambda\leq1$, $\mu$, $\nu$ are
positive constants. We use energy conservation equation and express
the density of MCG in this form
\begin{equation}\label{2x}
\rho_{mcg}=\bigg(A+\frac{c}{a^{3(1+\lambda)(1+\mu)}}\bigg)^{\frac{1}{1+\lambda}}
,
\end{equation}
where $c$ is constant of integration and $ A = \frac{\nu}{1+\mu}$ .
We start with the modified Friedmann equation of the form
\begin{equation}\label{14}
H^{2}=\frac{1}{3M_{p}^{2}}\bigg(\big(A+\rho_{\phi}^{(1+\lambda)(1+\mu)}\big)^{\frac{1}{1+\lambda}}+\rho_{\gamma}\bigg),
\end{equation}
this modification is possible only due to an extrapolation of
Eq.(\ref{2x}) so that
\begin{eqnarray}\label{15}
\rho_{mcg}=\big(A+\rho_{m}^{(1+\lambda)(1+\mu)}\big)^{\frac{1}{1+\lambda}}
\rightarrow\big(A+\rho_{\phi}^{(1+\lambda)(1+\mu)}\big)^{\frac{1}{1+\lambda}},
\end{eqnarray}
where $\rho_{m}$ denotes the matter energy density, and hence the
Friedmann equation takes the form
\begin{eqnarray}\label{16}
H^{2}=\frac{1}{3M_{p}^{2}}(A+\rho_{\phi}^{(1+\lambda)(1+\mu)})^{\frac{1}{1+\lambda}}
\sim
\frac{1}{3M_{p}^{2}}(A+V^{(1+\lambda)(1+\mu)})^{\frac{1}{1+\lambda}}.
\end{eqnarray}

\subsubsection{Weak Dissipative Regime}

For weak dissipative regime the temperature remains same as given in
GCG case. However, the slow roll parameters takes the following form
\begin{eqnarray}\nonumber
\epsilon & =&
\frac{M_{p}^{2}(1+\mu)V^{(1+\lambda)(1+\mu)-1}V'^{2}}{2(A
+V^{(1+\lambda)(1+\mu)})^{\frac{2+\lambda}{1+\lambda}}}
,\\\nonumber\eta
&=&\frac{M_{p}^{2}}{(A+V^{(1+\lambda)(1+\mu)})^{\frac{1}{1
+\lambda}}}\bigg(2V''+\frac{V'^{2}((1+\lambda)(1+\mu)-1)}{V}
\\\nonumber&&-\frac{V^{(1+\lambda)(1+\mu)-1}V'^{2}(1+\lambda)(1+\mu)}{(A+V^{(1+\lambda)(1+\mu)})}\bigg)
,\\\nonumber\beta &=&
M_{p}^{2}\bigg(\frac{4(A+V^{(1+\lambda)(1+\mu)})V''-3(1+\mu)
V'^{2}V^{(1+\lambda)(1+\mu)-1}}{6(A+V^{(1+\lambda)(1+\mu)})^{\frac{2+\lambda}{1+\lambda}}}\bigg).
\end{eqnarray}
The number of e-folds are obtained as
\begin{eqnarray}\nonumber
N=\frac{1}{M_{p}^{2}}\int_{\phi_{end}}^{\phi_{*}}\frac{(A+V^{(1+\lambda)(1+\mu)})^{\frac{1}{1+\lambda}}}{V'}d\phi.
\end{eqnarray}
Other perturbed parameters turns out to be
\begin{eqnarray}\label{18}
P_{R}&=&\bigg(\frac{81\pi
b^{2}}{12^{2}CV'^{2}}\bigg)^{\frac{1}{2}}\bigg(\frac{(A
+V^{(1+\lambda)(1+\mu)})^{\frac{1}{1+\lambda}}}{3M_{p}^{2}}\bigg)^{\frac{3}{2}}
, \\\label{23}
n_{s}-1&=&\frac{3M_{p}^{2}}{2(A+V^{(1+\lambda)(1+\mu)})^{\frac{1}{1+\lambda}}}\bigg(
\frac{V'^{2}((1+\lambda)(1+\mu)-1)}{V}\\\nonumber&&
-\frac{V^{(1+\lambda)(1+\mu)-1}V'^{2}}{(A+V^{(1+\lambda)(1+\mu)})}\big(\frac{3}{4}(1+\mu)+(1+\lambda)(1+\mu)\big)+2V''
\\\nonumber&&-\frac{3}{2}\big(\frac{4(A+V^{(1+\lambda)(1+\mu)})V''
-3(1+\mu)V'^{2}V^{(1+\lambda)(1+\mu)-1}}{6(A+V^{(1+\lambda)(1+\mu)})}\big)\bigg),
\\\label{20} r&=&
\frac{192G\sqrt{3CM_{p}^{2}}V'}{9b\pi^{\frac{3}{2}}(A+V^{(1+\lambda)(1+\mu)})^{\frac{1}{2(1+\lambda)}}}.
\end{eqnarray}
By using a quartic potential $r$ and $n_{s}$ are expressed as
function of $\phi$,
\begin{eqnarray}\label{23}
n_{s}-1&=&\frac{3M_{p}^{2}}{2(A+(0.25\lambda_{*}\phi^{4})^{(1+\lambda)(1+\mu)})^{\frac{1}{1+\lambda}}}\bigg(
\lambda_{*}\phi^{2}((1+\lambda)(1+\mu)-1)\\\nonumber&&
-\frac{(0.25\lambda_{*}\phi^{4})^{(1+\lambda)(1+\mu)-1}\lambda_{*}^{2}\phi^{6}}{(A+(0.25\lambda_{*}\phi^{4})^{(1+\lambda)(1+\mu)})}\big(\frac{3}{4}(1+\mu)+(1+\lambda)(1+\mu)\big)+6\lambda_{*}\phi^{2}
\\\nonumber&&-\frac{3}{12(A+(0.25\lambda_{*}\phi^{4})^{(1+\lambda)(1+\mu)})}\big(4(A
+(0.25\lambda_{*}\phi^{4})^{(1+\lambda)(1+\mu)})\lambda_{*}\phi^{2}\\\nonumber&&-3(1+\mu)\lambda_{*}^{2}\phi^{6}(0.25\lambda_{*}\phi^{4})^{(1+\lambda)(1+\mu)-1}\big)\bigg),
\\\label{20} r&=&
\frac{192G\sqrt{3CM_{p}^{2}}\lambda_{*}\phi^{3}}{9b\pi^{\frac{3}{2}}(A
+(0.25\lambda_{*}\phi^{4})^{(1+\lambda)(1+\mu)})^{\frac{1}{2(1+\lambda)}}}.
\end{eqnarray}
\begin{figure} \centering
\epsfig{file=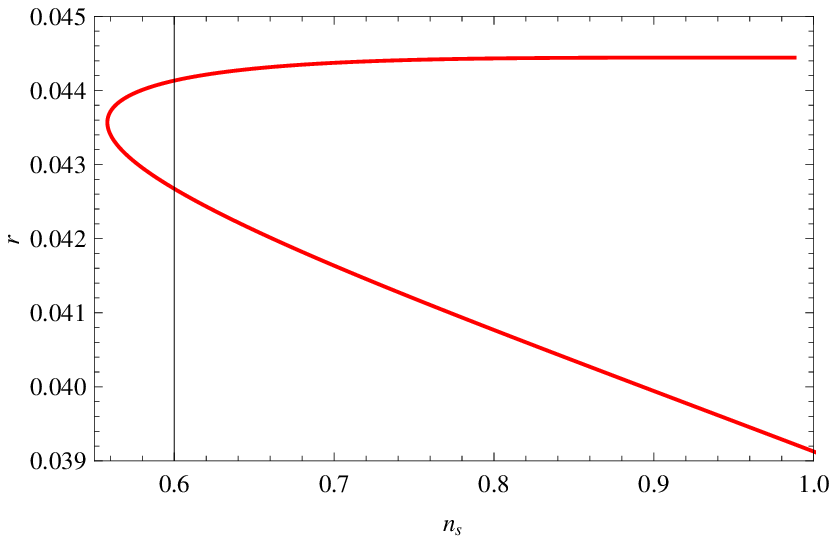,width=.45\linewidth}
\epsfig{file=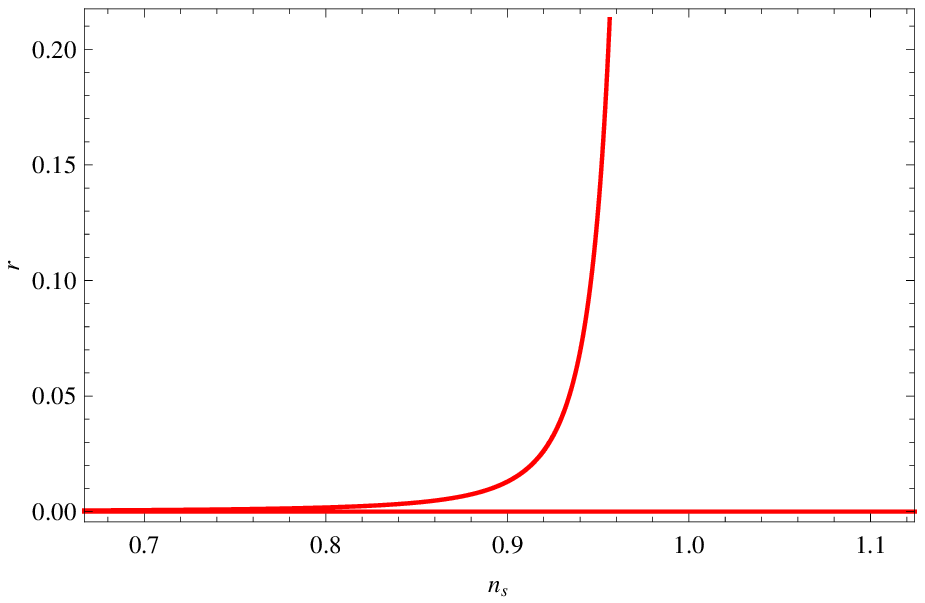,width=.45\linewidth}\caption{Plot of
tensor-scalar ratio $r$ versus scalar spectral index $n_{s}$ for MCG
model in weak dissipative regime (left panel) and strong dissipative
regime (right panel)}
\end{figure}

\subsubsection{Strong Dissipative Regime}

Here, we mention that the temperature remains same as obtained in
GCG case for strong regime. However, the slow roll parameters take
the form
\begin{eqnarray}\nonumber
\epsilon & =&
\frac{M_{p}^{2}(1+\mu)V^{(1+\lambda)(1+\mu)-1}V'^{2}}{2R(A+V^{(1+\lambda)(1+\mu)})^{\frac{2+\lambda}{1+\lambda}}}
,\\\nonumber\eta
&=&\frac{M_{p}^{2}}{R(A+V^{(1+\lambda)(1+\mu)})^{\frac{1}{1+\lambda}}}\bigg(2V''
+\frac{V'^{2}((1+\lambda)(1+\mu)-1)}{V}\\\nonumber&&-\frac{V^{(1+\lambda)(1+\mu)
-1}V'^{2}(1+\lambda)(1+\mu)}{(A+V^{(1+\lambda)(1+\mu)})}\bigg)
,\\\nonumber\beta &=&
\frac{1}{R}M_{p}^{2}\bigg(\frac{4(A+V^{(1+\lambda)(1+\mu)})V''-(1+\mu)V'^{2}
V^{(1+\lambda)(1+\mu)-1}}{10(A+V^{(1+\lambda)(1+\mu)})^{\frac{2+\lambda}{1+\lambda}}}\bigg).
\end{eqnarray}
The number of e-folds is given by
\begin{eqnarray}\nonumber
N=\frac{1}{M_{p}^{2}}\int_{\phi_{end}}^{\phi_{*}}\frac{(A+V^{(1+\lambda)(1+\mu)})^{\frac{1}{1+\lambda}}}{V'}Rd\phi.
\end{eqnarray}
Other perturbed quantities lead to
\begin{eqnarray}\label{18}
P_{R}&=&\big(\frac{\pi}{4}\big)^{\frac{1}{2}}\frac{b^{\frac{9}{5}}}{V'^{\frac{3}{5}}(4C)^{\frac{7}{10}}}\bigg(\frac{(A+V^{(1+\lambda)(1+\mu)})^{\frac{1}{1+\lambda}}}{3M_{p}^{2}}\bigg)^{\frac{9}{10}}
, \\\label{23}
n_{s}-1&=&\frac{3(4C)^{\frac{1}{5}}}{2b^{\frac{4}{5}}V'^{\frac{2}{5}}}\frac{(3M_{p}^{2})^{\frac{2}{5}}}{(A+V^{(1+\lambda)(1+\mu)})^{\frac{2}{5(1+\lambda)}}}\bigg(
\frac{V'^{2}((1+\lambda)(1+\mu)-1)}{V}\\\nonumber&&-\frac{V^{(1+\lambda)(1+\mu)-1}V'^{2}}{(A+V^{(1+\lambda)(1+\mu)})}\big(\frac{3}{4}(1+\mu)+(1+\lambda)(1+\mu)\big)+2V''
\\\nonumber&&-\frac{3}{2}\bigg(\frac{4(A+V^{(1+\lambda)(1+\mu)}V''-(1+\mu)V'^{2}V^{(1+\lambda)(1+\mu)-1})}{10(A+V^{(1+\lambda)(1+\mu)})}\bigg)\bigg)
, \\\label{24} r&=&
\frac{32G(4C)^{\frac{7}{10}}V'^{\frac{3}{5}}}{b^{\frac{9}{5}}\pi^{\frac{3}{2}}}\bigg(\frac{(A+V^{(1+\lambda)(1+\mu)})^{\frac{1}{1+\lambda}}}{3M_{p}^{2}}\bigg)^{\frac{1}{10}}
.
\end{eqnarray}
By putting the value of $V$ and $V'$,  we get $r$ and $n_{s}$ in
terms of $\phi$,
\begin{eqnarray}\nonumber
n_{s}-1&=&\frac{3(4C)^{\frac{1}{5}}}{2b^{\frac{4}{5}}(\lambda_{*}\phi^{3})^{\frac{2}{5}}}
\frac{(3M_{p}^{2})^{\frac{2}{5}}}{(A+(0.25\lambda_{*}\phi^{4})^{(1
+\lambda)(1+\mu)})^{\frac{2}{5(1+\lambda)}}}\bigg(
\lambda_{*}\phi^{2}((1+\lambda)(1+\mu)\\\nonumber&&-1)
-\frac{(0.25\lambda_{*}\phi^{4})^{(1+\lambda)(1+\mu)-1}\lambda_{*}^{2}\phi^{6}}{(A+(0.25\lambda_{*}\phi^{4})^{(1+\lambda)(1+\mu)})}\big(\frac{3}{4}(1+\mu)+(1+\lambda)(1+\mu)\big)
\\\nonumber&&-\frac{3}{20(A+(0.25\lambda_{*}\phi^{4})^{(1+\lambda)
(1+\mu)})}\big(4(A+(0.25\lambda_{*}\phi^{4})^{(1+\lambda)(1+\mu)})
\lambda_{*}\phi^{2}\\\nonumber&&-(1+\mu)\lambda_{*}^{2}\phi^{6}(0.25
\lambda_{*}\phi^{4})^{(1+\lambda)(1+\mu)-1}\big)+6\lambda_{*}\phi^{2}\bigg),\\\nonumber
r&=&
\frac{32G(4C)^{\frac{7}{10}}(\lambda_{*}\phi^{3})^{\frac{3}{5}}}
{b^{\frac{9}{5}}\pi^{\frac{3}{2}}}\bigg(\frac{(A+(0.25\lambda_{*}\phi^{4})^{(1+\lambda)
(1+\mu)})^{\frac{1}{1+\lambda}}}{3M_{p}^{2}}\bigg)^{\frac{1}{10}}.
\end{eqnarray}
It can be noted from the Figure \textbf{2} that plots of $r$ in
terms of $n_{s}$ for MCG models in weak and strong regime where $r$
and $n_{s}$ are expressed as function of $\phi$. The parameters
appearing in the model have following values $\lambda=1$, $\mu=0.5$,
$\lambda_{*}=10^{-3}$, $A=10^{-25}$, $b=25$. The range of tensor
scalar ratio is $r<0.045$, when spectral index is $0.6<n_{s}<1$, in
weak regime (left panel). However, we get $r<0.15$ for $0.7<n_{s}<1$
with $b=60$ for strong dissipative regime (right panel). The
observed range of $r$ and $n_{s}$ is compatible with data provided
by WMAP9 \cite{17x} and Planck \cite{25}.

\subsection{Generalized Cosmic Chaplygin Gas}

Gozalez Diaz \cite{11x} introduced the GCCG model, its equation of
state is given by
\begin{equation}\nonumber
P_{gccg}=- \rho^{-\lambda}[A+(\rho_{gccg}^{1+\lambda} -
A)^{-\omega}] ,
\end{equation}
where $A=\frac{D}{1+\omega}-1$ and $D$ can taken as positive or
negative value. $\lambda$ is a positive constant and $-l<\omega<0$,
$l>1$. If we take $\omega\rightarrow 0$ then this equation of state
reduces to GCG model. We obtain the energy density of GCCG  by
integrating energy conservation equation
\begin{equation}\label{3x}
\rho_{gccg} = \bigg[A +
\big(1+\frac{B}{a^{3(1+\lambda)(1+\omega)}}\big)^{\frac{1}{1+\omega}}\bigg]^{\frac{1}{1+\lambda}}
,
\end{equation}
The modified Friedmann equation in view of GCCG becomes
\begin{equation}\label{14}
H^{2}=\frac{1}{3M_{p}^{2}}\bigg(\big(A+(1+\rho_{\phi}^{(1+\lambda)(1
+\omega)})^{\frac{1}{1+\omega}})\big)^{\frac{1}{1+\lambda}}+\rho_{\gamma}\bigg).
\end{equation}
This modification is possible only due to an extrapolation of
Eq.(\ref{3x}) so that
\begin{eqnarray}\label{15}
\rho_{gccg}=\big(A+(1+\rho_{m}^{(1+\lambda)(1+\omega)})^{\frac{1}{1
+\omega}}\big)^{\frac{1}{1+\lambda}}\rightarrow\big(A+(1+\rho_{\phi}^{(1
+\lambda)(1+\omega)})^{\frac{1}{1+\omega}}\big)^{\frac{1}{1+\lambda}},
\end{eqnarray}
For this case,  the Friedmann equation takes the form
\begin{eqnarray}\label{16}
H^{2}&=&\frac{1}{3M_{p}^{2}}(A+(1+\rho_{\phi}^{(1+\lambda)
(1+\omega)})^{\frac{1}{1+\omega}})^{\frac{1}{1+\lambda}},\\\nonumber&
\sim&
\frac{1}{3M_{p}^{2}}(A+(1+V^{(1+\lambda)(1+\omega)})^{\frac{1}{1+\omega}})^{\frac{1}{1+\lambda}}.
\end{eqnarray}

\subsubsection{Weak Dissipative Regime}

In this regime, the slow roll parameters become
\begin{eqnarray}\nonumber
\epsilon & =&
\frac{M_{p}^{2}V^{(1+\lambda)(1+\omega)-1}(1+V^{(1+\lambda)
(1+\omega)})^{\frac{-\omega}{1+\omega}}V'^{2}}{2(A+(1+V^{(1+\lambda)
(1+\omega)})^{\frac{1}{1+\omega}})^{\frac{2+\lambda}{1+\lambda}}}
,\\\nonumber\eta
&=&\frac{M_{p}^{2}}{(A+(1+V^{(1+\lambda)(1+\omega)})^{\frac{1}{1+\omega}})^{\frac{1}
{1+\lambda}}}\bigg(2V''+\frac{V'^{2}((1+\lambda)(1+\omega)-1)}{V}\\\nonumber&&
-\frac{\omega(1+\lambda)V^{(1+\lambda)(1+\omega)-1}V'^{2}}{(1+V^{(1+\lambda)(1+\omega)})}
-\frac{(1+V^{(1+\lambda)(1+\omega)})^{\frac{-\omega}{1+\omega}}}{(A+(1+V^{(1+\lambda)
(1+\omega)})^{\frac{1}{1+\omega}})}V'^{2}\\\nonumber&&(1+\lambda)V^{(1+\lambda)(1+\omega)-1}\bigg)
,\\\nonumber\beta&=&
\bigg(\frac{4(A+(1+V^{(1+\lambda)(1+\omega)})^{\frac{1}{1+\omega}})V''-3V'^{2}V^{(1+\lambda)
(1+\omega)-1}(1+V^{(1+\lambda)(1+\omega)})^{\frac{-\omega}{1+\omega}}}{6(A+(1+V^{(1+\lambda)
(1+\omega)})^{\frac{1}{1+\omega}})^{\frac{2+\lambda}{1+\lambda}}}\bigg)\\\nonumber&\times&M_{p}^{2}.
\end{eqnarray}
By using Eq.(\ref{1n}), the number of e-folds is given as
\begin{eqnarray}\nonumber
N=\frac{1}{M_{p}^{2}}\int_{\phi_{end}}^{\phi_{*}}\frac{(A+(1+V^{(1+\lambda)(1
+\mu)})^{\frac{1}{1+\omega}})^{\frac{1}{1+\lambda}}}{V'}d\phi.
\end{eqnarray}
The perturbed parameters take the form
\begin{eqnarray}\label{18}
P_{R}&=&\bigg(\frac{81\pi
b^{2}}{12^{2}CV'^{2}}\bigg)^{\frac{1}{2}}\bigg(\frac{(A+(1
+V^{(1+\lambda)(1+\omega)})^{\frac{1}{1+\omega}})^{\frac{1}{1+\lambda}}}{3M_{p}^{2}}\bigg)^{\frac{3}{2}}
, \\\nonumber
n_{s}-1&=&\frac{3M_{p}^{2}}{2(A+(1+V^{(1+\lambda)(1
+\omega)})^{\frac{1}{1+\omega}})^{\frac{1}{1+\lambda}}}\bigg(\frac{V'^{2}((1
+\lambda)(1+\omega)-1)}{V}\\\nonumber&&-\frac{3}{12(A+(1+V^{(1+\lambda)(1
+\omega)})^{\frac{1}{1+\omega}})}\bigg(4V''(A+(1+V^{(1+\lambda)(1+\omega)})^{\frac{1}{1+\omega}})\\\nonumber&&
-3V'^{2}V^{(1+\lambda)(1+\omega)-1}(1+V^{(1+\lambda)(1+\omega)})^{\frac{-\omega}{1+\omega}}\bigg)+2V''\\\nonumber&&
-\frac{\omega(1+\lambda)V^{(1+\lambda)(1+\omega)-1}V'^{2}}{(1+V^{(1+\lambda)(1+\omega)})}
-\frac{(1+V^{(1+\lambda)(1+\omega)})^{\frac{-\omega}{1+\omega}}}{(A
+(1+V^{(1+\lambda)(1+\omega)})^{\frac{1}{1+\omega}})}V'^{2}\\\label{31}&&
(1+\lambda)V^{(1+\lambda)(1+\omega)-1}(\frac{3}{4}+(1+\lambda))\bigg)
, \\\label{20}
r&=&
\frac{192G\sqrt{3CM_{p}^{2}}V'}{9b\pi^{\frac{3}{2}}(A+(1+V^{(1+\lambda)(1+\mu)})^{\frac{1}{1+\omega}})^{\frac{1}{2(1+\lambda)}}}.
\end{eqnarray}
We can write $r$ and $n_{s}$ in terms of $\phi$ as follows
\begin{eqnarray}\nonumber
n_{s}-1&=&\frac{3M_{p}^{2}}{2(A+(1+(0.25\lambda_{*}\phi^{4})^{(1+\lambda)(1+\omega)})^{\frac{1}{1
+\omega}})^{\frac{1}{1+\lambda}}}\bigg(4((1+\lambda)(1+\omega)-1)\\\nonumber&&\lambda_{*}\phi^{2}
+6\lambda_{*}\phi^{2}-\frac{3}{12(A+(1+(0.25\lambda_{*}\phi^{4})^{(1+\lambda)(1+\omega)})^{\frac{1}{1
+\omega}})}\bigg(12\lambda_{*}\phi^{2}\\\nonumber&&(A+(1+(0.25\lambda_{*}\phi^{4})^{(1+\lambda)(1
+\omega)})^{\frac{1}{1+\omega}})-3\lambda_{*}^{2}\phi^{6}(0.25\lambda_{*}\phi^{4})^{(1+\lambda)(1
+\omega)-1}\\\nonumber&&(1+(0.25\lambda_{*}\phi^{4})^{(1+\lambda)(1+\omega)})^{\frac{-\omega}{1
+\omega}}\bigg)-\frac{(0.25\lambda_{*}\phi^{4})^{(1+\lambda)(1+\omega)-1}\lambda_{*}^{2}\phi^{6}}
{(1+(0.25\lambda_{*}\phi^{4})^{(1+\lambda)(1+\omega)})}
\\\nonumber&&\omega(1+\lambda)-\frac{(1+(0.25\lambda_{*}\phi^{4})^{(1
+\lambda)(1+\omega)})^{\frac{-\omega}{1+\omega}}}{(A+(1+(0.25\lambda_{*}\phi^{4})^{(1+\lambda)
(1+\omega)})^{\frac{1}{1+\omega}})}(\frac{3}{4}+(1+\lambda))\\\nonumber&&\lambda_{*}^{2}\phi^{6}(1
+\lambda)(0.25\lambda_{*}\phi^{4})^{(1+\lambda)(1+\omega)-1}\bigg)
, \\\nonumber
r&=&\frac{192G\sqrt{3CM_{p}^{2}}\lambda_{*}\phi^{3}}{9b\pi^{\frac{3}{2}}(A
+(1+(0.25\lambda_{*}\phi^{4})^{(1+\lambda)(1+\mu)})^{\frac{1}{1+\omega}})^{\frac{1}{2(1+\lambda)}}}.
\end{eqnarray}

\subsubsection{Strong Dissipative Regime}

For strong regime, the slow roll parameters takes the form
\begin{eqnarray}\nonumber
\epsilon & =&
\frac{M_{p}^{2}V^{(1+\lambda)(1+\omega)-1}(1+V^{(1+\lambda)(1+\omega)})^{\frac{-\omega}{1
+\omega}}V'^{2}}{2R(A+(1+V^{(1+\lambda)(1+\omega)})^{\frac{1}{1+\omega}})^{\frac{2+\lambda}{1+\lambda}}}
,\\\nonumber\eta
&=&\frac{M_{p}^{2}}{R(A+(1+V^{(1+\lambda)(1+\omega)})^{\frac{1}{1+\omega}})^{\frac{1}
{1+\lambda}}}\bigg(2V''+\frac{V'^{2}((1+\lambda)(1+\omega)-1)}{V}\\\nonumber&&
-\frac{\omega(1+\lambda)V^{(1+\lambda)(1+\omega)-1}V'^{2}}{(1+V^{(1+\lambda)(1+\omega)})}
-\frac{(1+V^{(1+\lambda)(1+\omega)})^{\frac{-\omega}{1+\omega}}}{(A+(1
+V^{(1+\lambda)(1+\omega)})^{\frac{1}{1+\omega}})}V'^{2}\\\nonumber&&(1+\lambda)V^{(1+\lambda)(1+\omega)-1}\bigg)
,\\\nonumber\beta &=&
\bigg(\frac{4(A+(1+V^{(1+\lambda)(1+\omega)})^{\frac{1}{1+\omega}})
-V'^{2}V^{(1+\lambda)(1+\omega)-1}(1+V^{(1+\lambda)(1+\omega)})^{\frac{-\omega}{1
+\omega}}}{10(A+(1+V^{(1+\lambda)(1+\omega)})^{\frac{1}{1+\omega}})^{\frac{2
+\lambda}{1+\lambda}}}\bigg)\\\nonumber&&\frac{1}{R}M_{p}^{2},
\end{eqnarray}
The number of e-folds leads to
\begin{eqnarray}\nonumber
N=\frac{1}{M_{p}^{2}}\int_{\phi_{end}}^{\phi_{*}}\frac{(A
+(1+V^{(1+\lambda)(1+\mu)})^{\frac{1}{1+\omega}})^{\frac{1}{1+\lambda}}}{V'}Rd\phi.
\end{eqnarray}
The corresponding perturbed quantities become
\begin{eqnarray}\label{18}
P_{R}&=&\big(\frac{\pi}{4}\big)^{\frac{1}{2}}\frac{b^{\frac{9}{5}}}{V'^{\frac{3}{5}}(4C)^{\frac{7}{10}}}
\bigg(\frac{(A+(1+V^{(1+\lambda)(1+\omega)})^{\frac{1}{1+\omega}})^{\frac{1}{1
+\lambda}}}{3M_{p}^{2}}\bigg)^{\frac{9}{10}}\\\nonumber
n_{s}-1&=&\frac{3(4C)^{\frac{1}{5}}}{2b^{\frac{4}{5}}V'^{\frac{2}{5}}}\frac{(3M_{p}^{2})^{\frac{2}{5}}}{(A
+(1+V^{(1+\lambda)(1+\omega)})^{\frac{1}{1+\omega}})^{\frac{2}{5(1+\lambda)}}}\bigg(\frac{V'^{2}((1
+\lambda)(1+\omega)-1)}{V}\\\nonumber&&-\frac{3}{20(A+(1+V^{(1+\lambda)(1+\omega)})^{\frac{1}{1
+\omega}})}\bigg(4V''(A+(1+V^{(1+\lambda)(1+\omega)})^{\frac{1}{1+\omega}})\\\nonumber&&-V'^{2}
V^{(1+\lambda)(1+\omega)-1}(1+V^{(1+\lambda)(1+\omega)})^{\frac{-\omega}{1+\omega}}\bigg)
+2V''\\\nonumber&&-\frac{\omega(1+\lambda)V^{(1+\lambda)(1+\omega)-1}V'^{2}}{(1+V^{(1+\lambda)(1+\omega)})}
-\frac{(1+V^{(1+\lambda)(1+\omega)})^{\frac{-\omega}{1+\omega}}}{(A+(1+V^{(1+\lambda)(1
+\omega)})^{\frac{1}{1+\omega}})}V'^{2}\\\label{31}&&(1+\lambda)V^{(1+\lambda)(1+\omega)
-1}(\frac{3}{4}+(1+\lambda))\bigg),
\end{eqnarray}
\begin{equation}\label{24}
r =
\frac{32G(4C)^{\frac{7}{10}}V'^{\frac{3}{5}}}{b^{\frac{9}{5}}\pi^{\frac{3}{2}}}
\bigg(\frac{(A+(1+V^{(1+\lambda)(1+\omega)})^{\frac{1}{1+\omega}})^{\frac{1}{1+\lambda}}}{3M_{p}^{2}}\bigg)^{\frac{1}{10}}
.
\end{equation}
However, $r$ and $n_{s}$ as function of $\phi$ are
\begin{eqnarray}\nonumber
n_{s}-1&=&\frac{3(4C)^{\frac{1}{5}}}{2b^{\frac{4}{5}}(\lambda_{*}\phi^{3})^{\frac{2}{5}}}
\frac{(3M_{p}^{2})^{\frac{2}{5}}}{(A+(1+(0.25\lambda_{*}\phi^{4})^{(1
+\lambda)(1+\omega)})^{\frac{1}{1+\omega}})^{\frac{2}{5(1+\lambda)}}}\bigg(4\lambda_{*}\phi^{2}\\\nonumber&&((1+\lambda)
(1+\omega)-1)-\frac{3}{20(A+(1+(0.25\lambda_{*}\phi^{4})^{(1+\lambda)(1+\omega)})^{\frac{1}{1
+\omega}})}\bigg(12\\\nonumber&&\lambda_{*}\phi^{2}(A+(1+(0.25\lambda_{*}\phi^{4})^{(1+\lambda)(1
+\omega)})^{\frac{1}{1+\omega}})-(0.25\lambda_{*}\phi^{4})^{(1+\lambda)(1+\omega)-1}\\\nonumber&&
\lambda_{*}^{2}\phi^{6}(1+(0.25\lambda_{*}\phi^{4})^{(1+\lambda)(1+\omega)})^{\frac{-\omega}{1
+\omega}}\bigg)-\frac{(0.25\lambda_{*}\phi^{4})^{(1+\lambda)(1+\omega)-1}\lambda_{*}^{2}\phi^{6}}{(1
+(0.25\lambda_{*}\phi^{4})^{(1+\lambda)(1+\omega)})}
\\\nonumber&&\omega(1+\lambda)-\frac{(1+(0.25\lambda_{*}\phi^{4})^{(1
+\lambda)(1+\omega)})^{\frac{-\omega}{1+\omega}}}{(A+(1+(0.25\lambda_{*}\phi^{4})^{(1
+\lambda)(1+\omega)})^{\frac{1}{1+\omega}})}(\frac{3}{4}+(1+\lambda))\\\nonumber&&
\lambda_{*}^{2}\phi^{6}(1+\lambda)(0.25\lambda_{*}\phi^{4})^{(1+\lambda)(1+\omega)-1}+6\lambda_{*}\phi^{2}\bigg)
,\\\nonumber r&=&
\frac{32G(4C)^{\frac{7}{10}}(\lambda_{*}\phi^{3})^{\frac{3}{5}}}
{b^{\frac{9}{5}}\pi^{\frac{3}{2}}}\bigg(\frac{(A+(1+(1+(0.25
\lambda_{*}\phi^{4})^{(1+\lambda)(1+\omega)})^{\frac{1}{1+\omega}})^{\frac{1}{1+\lambda}}}{3M_{p}^{2}}\bigg)^{\frac{1}{10}}.
\end{eqnarray}
\begin{figure} \centering
\epsfig{file=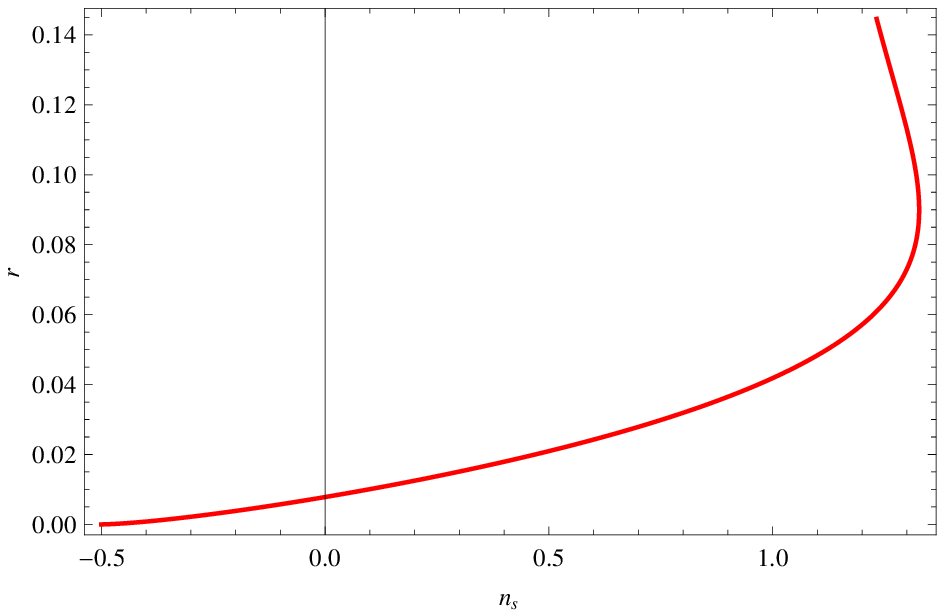,width=.45\linewidth}
\epsfig{file=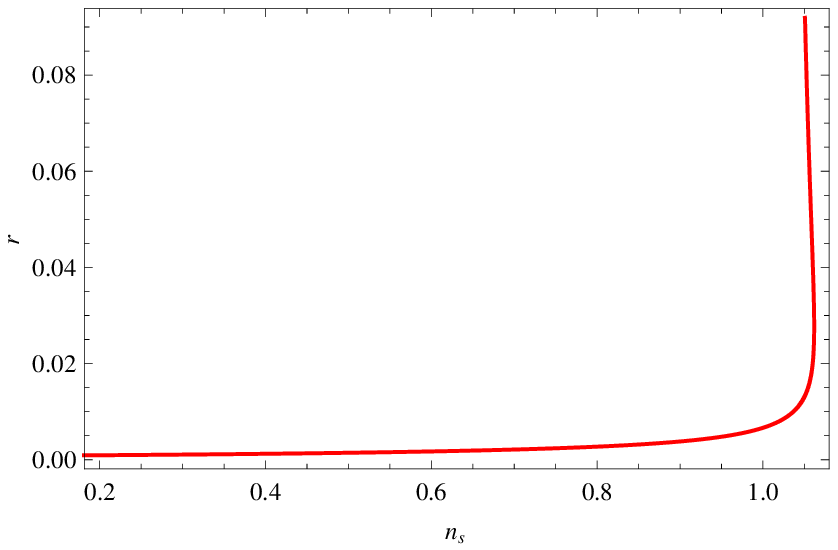,width=.45\linewidth}\caption{Plot of
tensor-scalar ratio $r$ versus scalar spectral index $n_{s}$ for
GCCG model in weak dissipative regime (left panel) and strong
dissipative regime (right panel)}
\end{figure}

The plots of $r$ versus $n_{s}$ for GCCG models in weak and strong
regime are shown in Figure \textbf{3}. The constant parameters are
$\lambda=1$, $\omega=-0.5$, $M_{p}=1$, $\lambda_{*}=10^{-2}$,
$A=10^{-5}$, $b=30$. In weak regime (left panel), the tensor scalar
ratio is confined to $r<0.12$ when spectral index is $n_{s}<1$. In
strong regime, we get $r<0.08$ for $b=80,~0.2<n_{s}<1$. These values
shows that GCCG model is compatible with data provided by WMAP$9$
and Planck \cite{17x,25}.

\section{Conclusions}

Warm inflation presents a compelling solution for the main problem
of the inflationary theory that how this inflationary period will
come to an end. In this type of models, radiations are produced
during inflation, and a dissipative coefficient is introduced. This
is the reason, we have investigated the warm inflationary scenario
inspired with quartic form of potential
$V=\frac{\lambda_{*}\phi^{4}}{4}$ and well-known form of dissipative
coefficient $\Gamma\propto T$. In order to find the consistency of
the results, we have assumed various well-known chaplygin gas models
such as GCG, MCG and GCCG. Also, we have considered that this
universe is filled with radiation and standard scalar field and
accordingly Friedmann equations are modified. Under slow roll
approximation, we have investigated inflationary parameters such as
number of e-folds, scalar spectrum, scaler spectral index, and
tensor to scalar ratio both in weak and strong dissipative regimes.

To analyze our results, we have plotted the graphs between tensor to
scalar ratio $r$ and scalar spectral index $n_{s}$ for each model in
weak (where $\Gamma\ll3H$) and strong (where $\Gamma\gg 3H$)
dissipative regimes. For GCG model, it is found that in weak
dissipative regime with $0.4<n_{s}<1$, we have $r<0.006$, and in
strong dissipative regime, $r=0.05$ at $n_{s}=0.96$ (referred as
Figure \textbf{1}). In MCG model, spectral index lies between
$0.6<n_{s}<1$, the range of tensor scalar ratio is $r<0.045$ in weak
regime. However, in strong regime, we have obtained the range
$r<0.15$ for $0.7<n_{s}<1$. In GCCG model, for the weak regime when
spectral index is $n_{s}<1$, the tensor scalar ratio is confined to
$r<0.12$. But, in strong regime for $b=80$, $0.2<n_{s}<1$, we get
$r<0.08$.

In addition, WMAP$9$ provides the value of tensor scalar ratio
$r<0.13$ and spectral index is measured to be $n_{s}=0.972\pm
0.013$, according to Planck data $r<0.11$ and $n_{s}=0.968\pm
0.006$. We have concluded with good remark that the obtain
range/values of $r$ corresponding to well-settled $n_{s}$ are well
supported to WMAP9 \cite{17x} and Planck data \cite{25} in all
models of CG models.


\begin{thebibliography}{43}


\bibitem{1} Starobinsky, A.A.: Phys. Lett. B
\textbf{91}(1980); Guth, A.: Phys. Rev. D \textbf{23}(1981)347.

\bibitem{2} Gold, B. et al.: Astrophy. J. Suppl.
\textbf{192}(2011)15.

\bibitem{3}  Berera, A.: Phys. Rev. Lett. \textbf{75}(1995)3218.

\bibitem{3t} Berera, A.: Phys. Rev. D \textbf{55}(1995)3346.

\bibitem{36}  Hall, L.M.H., Moss, I.G., Berera, A.: Phys. Rev.
 D \textbf{69}(2004)083525.

\bibitem{37} Berera, A.: Phys. Rev. D \textbf{54}(1996)2519.

\bibitem{41} Moss, I.G.: Phys. Lett. B \textbf{154}(1985)120.

\bibitem{42}  Berera, A., Fang, L.Z.: Phys. Rev. Lett. \textbf{74}(1995)1912.

\bibitem{43} Berera, A.: Nucl. Phys. B \textbf{585}(2000)666.

\bibitem{5t} Pich, A.: arXiv:0705.4264[hep-ph].

\bibitem{11} Monerat, G.A. et al.: Phys. Rev. D \textbf{76}(2007)02017.

\bibitem{4t}  Antonella cid, M., Del Campo, S., Herrera, R.: JCAP \textbf{0710}(2007)005.

\bibitem{9}  Del Campo, S., Herrera, R.: Phys. Lett. B \textbf{660}(2008)282.

\bibitem{2t}  Setare, M.R., Kamali, V.: JCAP \textbf{08}(2012).

\bibitem{4t}  Antonella cid, M., Del Campo, S., Herrera, R.: JCAP \textbf{0710}(2007)005.

\bibitem{6} Bastero-Gil, M., Berera, A., Ramos, R.O., Rosa, J.G.: JCAP \textbf{1301}(2013)016.

\bibitem{1t} Setare, M.R., Kamali, V.: Phys. Rev. D \textbf{87}(2013)083524.

\bibitem{9t}  Herrera, R., Olivares, M., Videla, N.: Eur. Phys. J. C \textbf{73}(2013)2295.

\bibitem{10} Herrera, R., Olivares, M., Videla, N.: Phys. Rev. D \textbf{88}(2013)063535.

\bibitem{10t} Herrera, R., Olivares, M., Videla, N.: Mod. Phys. D \textbf{23}(2014)1450080.


\bibitem{5} Bastero-Gil, M., Berera, A., Ramos, R.O., Rosa, J.G.:
JCAP \textbf{1410}(2014)10053.

\bibitem{8}  Sharif, M., Saleem, R.: Eur. Phys. J. C
\textbf{74}(2014).



\bibitem{3t} Setare, M.R., Kamali, V.: Class. Quantum Grav.
\textbf{32}(2015)235005.


\bibitem{7}  Setare, M.R., Kamali, V.: Int. J. Theor. Phys.
\textbf{55}(2016)103.



\bibitem{4} Panotopoulos, G., Videla, N.: Eur. Phys. J. C \textbf{75} (2015)

\bibitem{12} Zhang, Y.: JCAP \textbf{0903}(2009)023.

\bibitem{15}  Bastero-Gil, M., Berera, A., Ramos, R.O.: JCAP \textbf{1107}(2011)030.

\bibitem{13} Berera, A., Gleiser, M., Ramos, R.O.: Phys. Rev. D \textbf{58}(1998)123508.

\bibitem{14} Yokoyama, J., Linde, A.: Phys. Rev. D \textbf{60}(1999)083500.

\bibitem{8x} Bento, M.S., Bertolami, O., Sen, A.: Phys. Rev.
 D \textbf{66}(2002)043507.

\bibitem{9x}  Kamenshchik, A., Moschell, U. and Pasquier, V.: Phys. Lett. B \textbf{511}(2001)265.

\bibitem{17x} Ade, P.A.R., et al.: Astron. Astro phys. A \textbf{16}(2014)571.

\bibitem{25} Hinshaw, G., et al.: Astrophys. J. Suppl. \textbf{208}(2013)19.

\bibitem{10x}  Benaoum, H.B., Debnath, U., Banerjee, A. and Chakraborty, S.: Class. Quantum Grav. \textbf{21}(2011)5609.

\bibitem{11x} Gonzalez-Diaz, P.F.:  Phys. Rev. D \textbf{68}(2003)021303.


\end{thebibliography}
\end{document}